\documentclass[sigconf]{acmart}

\AtBeginDocument{%
  }

\usepackage{subfigure}
\usepackage{multirow}
\usepackage{tabularx}
\usepackage{makecell}
\pdfoutput=1

\renewcommand\footnotetextcopyrightpermission[1]{}
\setcopyright{none}
\renewcommand\footnotetextcopyrightpermission[1]{}
\setcopyright{none}
\makeatletter
\renewcommand\@formatdoi[1]{\ignorespaces}
\makeatother

\begin{document}

\title{SmartX Intelligent Sec: A Security Framework Based on Machine Learning and eBPF/XDP}

\author{Talaya Farasat}

\affiliation{%
	\institution{University of Passau}
	\streetaddress{Passau, Germany}
	\city{Passau}
	\country{Germany}}
	
\author{JongWon Kim}

\affiliation{%
	\institution{GIST}
	\streetaddress{Gwangju, South Korea}
	\city{Gwangju}
	\country{South Korea}}

\author{Joachim Posegga}

\affiliation{%
	\institution{University of Passau}
	\streetaddress{Passau, Germany}
	\city{Passau}
	\country{Germany}}

\renewcommand{\shortauthors}{Farasat et al. }

\begin{abstract}
 Information and Communication Technologies (ICT) infrastructures are becoming increasingly complex day by day, facing numerous challenges to support the latest networking paradigms. Security is undeniably a critical component for the effective functioning of these advanced ICT infrastructures. By considering the current network security challenges, we propose SmartX Intelligent Sec, an innovative intelligent security framework. SmartX Intelligent Sec leverages a combination of the lightweight extended Berkeley Packet Filter/eXpress Data Path (eBPF/XDP) for efficient network packet capturing and filtering malicious network traffic, and a Bidirectional Long Short-Term Memory (BiLSTM) classifier for network threat detection. Our real-time prototype demonstrates that SmartX Intelligent Sec offers comprehensive automation features, enabling continuous network packet capturing, effective network threat detection, and efficient filtering of malicious network traffic. This framework ensures enhanced security and operational efficiency for modern ICT infrastructures.
\end{abstract}

\pagestyle{plain}



\maketitle

\section{Introduction}
Network security remains a significant challenge, especially with the increasing sophistication of cyber threats like Distributed Denial of Service (DDoS) attacks. These attacks can severely disrupt critical infrastructure, leading to extensive financial losses and operational downtime. Traditional security measures often fall short in countering the volume and complexity of modern threats, necessitating the adoption of more advanced solutions.

Currently, eBPF/XDP (a light-weight, powerful virtual machine inside linux kernel) has attracted significant interest in the
research community and has been used for a variety of use cases, e.g, network virtualization, load balancing, and network packet filtering/network threat mitigation. 

Meanwhile, machine learning also offers powerful tools for enhancing network security that far exceed traditional methods in terms of speed, accuracy, and adaptability. As cyber threats become more sophisticated, the usage of machine learning into network security will continue to play a crucial role in detecting network threats and identifying various network anomalies.

In this paper, we propose SmartX Intelligent Sec, an innovative intelligent security framework leveraging powerful, lightweight eBPF/XDP tool in conjunction with latest machine learning/deep learning techniques. SmartX Intelligent Sec employs eBPF/XDP for continuous network packet capturing and filtering of malicious network traffic and integrates with BiLSTM for detecting network threats. Overall, this combination (eBPF/XDP with BiLSTM) allows SmartX Intelligent Sec to provide comprehensive automation features for continuous network packet capturing, network threat detection, and filtering malicious network traffic.

SmartX Intelligent Sec has been evaluated both in terms of network threat detection accuracy and the efficacy  of filtering malicious network traffic. SmartX Intelligent Sec, BiLSTM-based network threat detection module demonstrates outstanding performance, with an accuracy of 99.3\%, F-score of 99.3\%, and ROC-AUC of 99.9\%. Furthermore, our real-time SmartX Intelligent Sec prototype demonstrates that eBPF/XDP efficiently filters/drops 2,295,337 malicious network packets (in 15 sec. of DDoS attack), affirming its capability in actively mitigating networks threats.
\section{Related Work}
Previous studies explore eBPF-based network monitoring systems and ideas about integration of machine learning techniques with eBPF/XDP. The significant performance improvement with eBPF/XDP for fast network packet processing and saving resources as compared to traditional methods has already been highlighted \cite{xdp, xdp2, xdp3, xdp4}. Ben-Yair et al. \cite{ben} propose to use machine learning with eBPF for detecting anomalies. Similarly in \cite{bachl, naive, anand, syn, syn2, compart},  simple machine learning models are used in conjunction with eBPF based technology for the purpose of detecting various network intrusions. In \cite{thesis, young, usenix}, eBPF is used to develop solutions against Denial-of-Service attacks, but they do not use machine learning. 

After integrating simple machine learning models with eBPF/XDP, researchers also focus on utilizing complex and more effective deep learning models with eBPF and XDP. Zhang et al. \cite{bpflstm} deploys eBPF in conjunction with LSTM for forecasting network situations. Sakuraba et al. \cite{sakuraba} use eBPF with LSTM specifically for monitoring routers. Kostopoulos \cite{kos} employs eBPF/XDP with simple neural networks for detection and prevention from network attacks. Recently, Karma IDS \cite{karmaIDS} has been introduced, leveraging eBPF for packet capturing and LSTM for network threat detection. Our work is closely related to Karma IDS, which employs eBPF solely for packet capturing. We use eBPF/XDP, allowing our system not only to capture packets but also to filter out malicious packets. By combining eBPF and XDP, SmartX Intelligent Sec can not only capture network packets but also filter and block malicious network traffic directly. This dual approach provides security by mitigating threats in real-time and improves performance by reducing the overhead associated with traditional packet processing methods. Moreover, instead of using LSTM, we evaluate its advanced version, BiLSTM for netwok threat detection. It is noteworthy our BiLSTM-based network threat detection module achieves an accuracy of 99.3\% (better than Karma IDS). As BiLSTM trains in both forward and backward directions, indeed it can be a better option than a simple LSTM or with simple neural networks.

\section{SmartX Intelligent Sec Design}The design of SmartX Intelligent Sec is illustrated in Figure 1. Initially, the eBPF/XDP tool captures the network packets. These network packets are then analyzed by the BiLSTM-based machine learning tool to detect network threats. If the network traffic is identified as malicious, the eBPF/XDP tool filters/drops the malicious network packets, and the connection with that malicious IP is terminated. Conversely, if the network traffic is deemed normal, the connectivity is maintained.

\begin{figure}[tp]
	\centering
	\includegraphics[width=70mm, height=3.8cm]{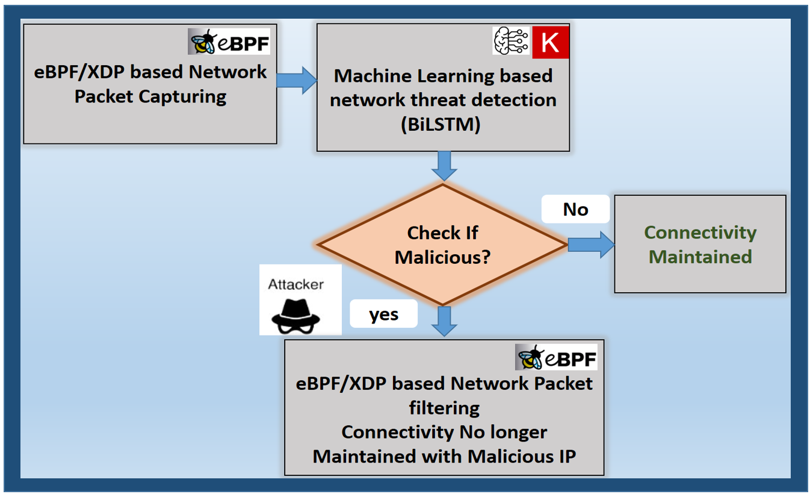}
	\caption{SmartX Intelligent Sec Design}

\end{figure}

\begin{figure}[tp]
	\centering
	\includegraphics[width=70mm, height=3.8cm]{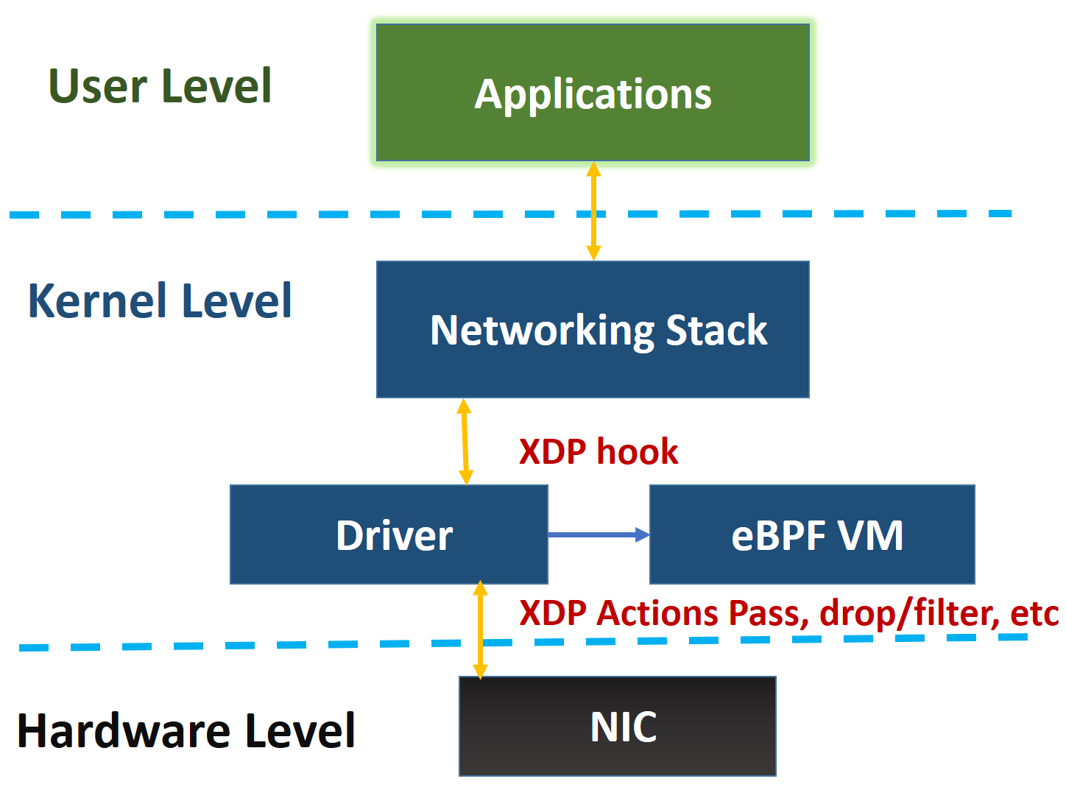}
	\caption{Linux network stack with eBPF/XDP}

\end{figure}

\subsection{Network Packet Capturing} Linux eBPF is a light-weight, powerful virtual machine that provides a set of libraries which allows dynamic injection of codes from user space into various kernel events, see Figure 2. Meanwhile, XDP provides special hooks for eBPF kernel programs to efficiently pass, drop/filter, and redirect network packets received at networking ports \cite{talia}. eBPF/XDP programs can be attached to the network driver with three different points, i.e., skb (generic), native, and hw (program offloaded to the hardware). The fastest one, i.e., hw or offload mode allows the XDP program runs on the NIC itself. However, for this mode, SmartNIC is required. In native mode, the XDP hook is called in the driver (driver support needed) before the kernel allocates the socket buffer. In skb or generic mode, XDP hook is called after the packet DMA and socket buffer allocation. So, in this case, the processing performance is significantly lower than in other modes. We configure the XDP tools \cite{bpf} and use specifically  xdp-dump (use native mode) for network packet capturing.

\subsection{Network Threat Detection}

For network threat detection, while we primarily focus on advanced deep learning techniques such as LSTM and BiLSTM, we first validate various simple machine learning algorithms to verify their performance accuracies.. So, we train various machine learning algorithms and evaluate their performance. The process begins with the selection of a suitable dataset, which is crucial for training machine learning models. In this section, first, we highlight our dataset and its key features. Next, we provide an overview of the different machine learning algorithms used in our study, explaining our selected corresponding hyper-parameters. Finally, we evaluate the performance of these algorithms, comparing metrics such as Accuracy, F-Score, and ROC to determine the best-performing model for network threat detection.

\begin{table*}	
	\renewcommand{\arraystretch}{0.4}
	\begin{tabular}{|p{2.8cm}|c|c|c|}
		\hline
		{\textbf{Features}} & {\textbf{Description}}\\
		
		\hline
		Flow Duration & \makecell{Length of connection in seconds} \\ \hline
		Fwd Pkt Len Max& \makecell{Maximum size of packets in the forward direction}\\ \hline
		Fwd Pkt Len Mean& \makecell{Average size of packets in the forward direction} \\ \hline
		Fwd Pkt Len Std& \makecell{Standard Deviation of the size of the packets in the \\ forward direction}  \\ \hline
		Flow IAT Mean & \makecell{Mean inter-arrival time between packet flows} \\ \hline
		Flow IAT Max & \makecell{Maximum inter-arrival time between packet flows} \\ \hline
		Fwd IAT Mean & \makecell{Forward inter-arrival time, the time between two packets \\sent forward direction Mean} \\ \hline
		
		Fwd IAT Tot & \makecell{Forward inter-arrival time, the time between two packets \\ sent forward direction Total} \\ \hline
		Fwd IAT Std & \makecell{Forward inter-arrival time, the time between two packets \\ sent forward direction Standard deviation} \\ \hline
		
		Fwd Header Len & \makecell{Total bytes used for a header in the forward direction} \\ \hline
		Pkt Len Min & \makecell{Minimum length of the packets flow} \\ \hline
		Pkt Len Var & \makecell{Minimum inter-arrival time of packet} \\ \hline
		SYN Flag Cnt & \makecell{Number of packets with SYN Flag} \\ \hline
		Fwd Seg Size Avg & \makecell{Average segment size in forward direction} \\ \hline
		Subflow Fwd Pkts & \makecell{Average number of packets in sub-flow in a forward direction} \\ \hline
		Active Max & \makecell{The amount of time a flow was active before becoming idle Max} \\ \hline
		Idle Std & \makecell{The amount of time a flow was idle before becoming active \\Standard deviation} \\ \hline
		Idle Mean & \makecell{The amount of time a flow was idle before becoming active Mean} \\ \hline
		Idle Min & \makecell{The amount of time a flow was idle before becoming active Min} \\ \hline
		Idle Max & \makecell{The amount of time a flow was idle before becoming active Max} \\ \hline		
	\end{tabular}
	\caption{Features used in training of Machine Learning Algorithms}
	
\end{table*}

\subsubsection{Dataset}We utilize our dataset available here \cite{dataset, talia2}. We configure CICFlowMeter-V4.0 \cite{CIC} to convert all .pcap files, encompassing both normal and malicious network traffic, into network flows. CICFlowMeter-V4.0 generates over 80 network traffic features. We use SelectKBest algorithm from scikit-learn \cite{scikit}, selecting the top 20 features, see Table 1. Our dataset comprises a total of 2.5 million network flows, balanced with 125,000 normal flows and 125,000 malicious flows. We frame this problem as a binary classification problem and use two labels, i.e., 0 for normal and 1 for malicious. We use Python scikit-learn train\_test\_split to create train and test datasets and use 70\% data for training and 30\% data for testing.

\subsubsection{Machine learning Algorithms}
 We develop various machine learning models using the Python scikit-learn library, including Support Vector Machine (SVM), Gaussian Naive Bayes (GNB), Logistic Regression (LR), Decision Tree, Extra-Trees, and Multilayer Perceptron (MLP). We implement Long Short-Term Memory (LSTM) and Bidirectional LSTM (biLSTM) by using the Python Keras framework. These models are chosen due to their proven effectiveness in addressing network threat detection challenges, as evidenced by prior research \cite{svm, gnb, logistic, DEC, et, mlp, lstm, bilstm, talia4, talia5, talia6, talia7}.  

\textbf{SVM} works by finding the optimal hyperplane that maximizes the margin between two classes of data. This ensures the greatest separation between the classes. SVM can be utilized with various kernel functions, but we choose a linear kernel for its simplicity and speed in binary classification tasks. All other parameters are kept at their default values.

\textbf{GNB} is a probabilistic algorithm based on Bayes' Theorem, which updates prior probabilities into posterior probabilities by incorporating the information, called likelihoods provided by the observed data. In this paper, we consider one of the extensions of naive bayes which is called Gaussian Naive Bayes which follows the Gaussian distribution.We implement a binary GNB classifier with all default parameters.

\textbf{LR} assigns data points to discrete classes by transforming its output with the logistic sigmoid function returning a probability value that can be mapped to discrete classes. We implement our binomial logistic regression, setting the solver to 'liblinear' while keeping all other parameters at their default values. 

\textbf{Decision Tree} is a flowchart-like structure with a root node, internal nodes, branches, and leaf nodes. The root node represents the entire dataset, while each internal node represents an input feature to perform a test. Branches emanating from a node represent the outcome of the test, that is, all possible feature values. Each leaf node represents a class label or a class probability distribution. A decision tree is learned by splitting the dataset into smaller subsets and tree is built by recursively splitting each derived subset until further splitting does not improve predictions. We create our Decision Tree model as a binary classifier, setting the max\_depth parameter to 3 while keeping all other parameters at their default values.

\textbf{Extra Trees} is an ensemble machine learning algorithm that combines the results of multiple de-correlated decision trees collected in a “forest” to output its classification results. We build our Extra Trees model as a binary classifier, setting max\_depth to 3 while keeping all other parameters at their default values.

\textbf{MLP} is a class of feedforward artificial neural networks that maps input data sets to a set of appropriate outputs. It is the most basic type of neural network. As scikit-learn is also capable of basic deep learning modeling, we use this library to generate our MLP model. We use all default parameters except max\_iter which we set as 300.

\textbf{LSTM} employs a special form of RNNs. Traditional RNNs are designed to store inputs in order to predict the outputs. But they are not performed well when several discrete time lags occur between the previous inputs and the present targets. To overcome this limitation, LSTM is capable of connecting time intervals to form a continuous memory. We use the Python Keras framework (backend Tensorflow) \cite{keras} to generate the LSTM model as a binary classifier. Our LSTM model contains an input layer, three hidden layers with 50 LSTM cells, a dropout layer with 20\% dropout rate, and an output layer with a single node. We use Adam optimizer and he model is trained for 50 epochs with the default learning rate = 0.001. 

\textbf{BiLSTM} in contrast with LSTM, processes data in both forward and backward directions using two separate hidden layers. These hidden layers are connected to the same output layer, enabling the model to traverse the input data twice. Bidirectional networks have been shown to outperform unidirectional LSTMs in network threat detection and various other fields \cite{bilstm}. We implement our BiLSTM model using the Python Keras framework with TensorFlow as the backend. The model features an input layer, a hidden layer with 64 LSTM units (BiLSTM creates from these LSTMs will be concatenated), and an output layer for binary classification. All other parameters are consistent with those used in the LSTM model.

\textbf{Performance Evaluation}: We evaluate the performance of our machine learning models based on Accuracy, F-score, and Receiver Operating Characteristic (ROC) metrics, see Table 2.
True positive (TP) and true negative (TN) are the samples that are correctly classified. False negative (FN) and false positive (FP) are  misclassified samples. Accuracy is the ratio of the number of correct predictions over the entire dataset. The F-score takes both false positives and false negatives into account. It is the weighted average harmonic value of precision and recall. Precision is the rate of true positives within all positives. Recall, also called sensitivity, is a measurement for the rate of positives that were correctly identified in comparison to the total number of actual positives. The ROC curve is the graphical representation of the true positive rate (TPR) against the false positive rate (FPR) rate at different classification thresholds.
 
\begin{equation}
	\label{eq:sedov}
	Accuracy= \frac{TP+TN}{TP+TN+FP+FN}\\
\end{equation}
\begin{equation}
	\label{eq:sedov}
	F-Score=\frac{Precision*Recall}{Precision+Recall}
\end{equation}
where,
\begin{equation}
	\label{eq:sedov}
	Precision=\frac{TP}{TP+FP}
\end{equation}
\begin{equation}
	\label{eq:sedov}
	Recall=\frac{TP}{TP+FN}
\end{equation}

Our results indicate that deep learning techniques, such as LSTM and BiLSTM, outperform traditional machine learning algorithms.  BiLSTM achieves the highest performance among all other machine learning algorithms, with an accuracy of 99.3\%, an F-score of 99.3\%, and an ROC of 99.9\%. Consequently, we choose BiLSTM for our real-time SmartX Intelligent Sec prototype.

Additionally, we measure the training times of two deep learning models (LSTM and BiLSTM) on both CPU and GPU, highlighted in Table 3. Our results indicate that training on an NVIDIA Tesla T4 GPU is significantly more efficient compared to an Intel® Xeon® D-2183IT CPU. We plan to implement real-time or incremental training in the future. If real-time training will be considered in future, we note that BiLSTM will require more time than LSTM but offers superior performance. This makes it crucial to decide whether to prioritize time efficiency or performance.
\begin{table}	
	\renewcommand{\arraystretch}{0.7}
	\begin{tabular}{|p{4.0cm}|c|c|c|}
		\hline
		{\textbf{Models}} & {\textbf{Accuracy}} & {\textbf{F-Score}}  & {\textbf{ROC}} \\

		\hline
		SVM & 83.06&85.3&94.4\\ \hline
		GNB& 81.9&84.3&83.0 \\ \hline
		LR&75.6&72.7&88.3  \\ \hline
		Decision Tree &82.3&84.8&92.8 \\ \hline
		Extra Trees & 86.4& 87.2&95.5 \\ \hline
		MLP &93.3&93.6&96.6\\ \hline
		LSTM (with GPU) &95.8&95.9&98.9 \\ \hline
		BiLSTM (with GPU) &99.3 &99.3 &99.9\\ \hline

	\end{tabular}
	\caption{Performance Evaluation of Machine Learning Algorithms}
	
\end{table}	

\begin{figure}[tp]
	\centering
	\includegraphics[width=85mm, height=3.3cm]{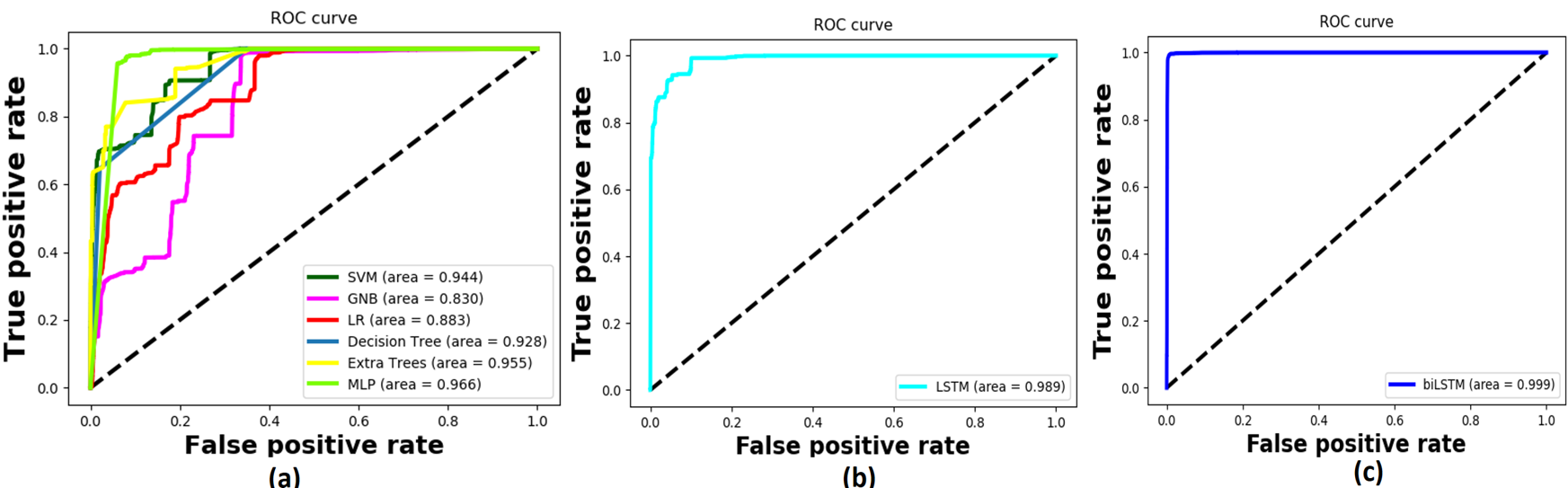}
	\caption{ROC result of different classifiers}

\end{figure}

\begin{table}	
	\renewcommand{\arraystretch}{0.8}
	\begin{tabular}{|p{4.0cm}|c|c|c|}
		\hline
		{} & {\textbf{LSTM}} &{\textbf{BiLSTM}} \\

		\hline
		CPU (Intel® Xeon® D-2183IT)& 5 min. 45 sec. &13 min. 33 sec.  \\ \hline
		GPU (Tesla T4)& 1 min. 9 sec.&2 min. 43 sec. \\ \hline

	\end{tabular}
	\caption{Time comparison between CPU and GPU}
	
\end{table}	
\begin{figure}[tp]
	\centering
	\includegraphics[width=90mm, height=4cm]{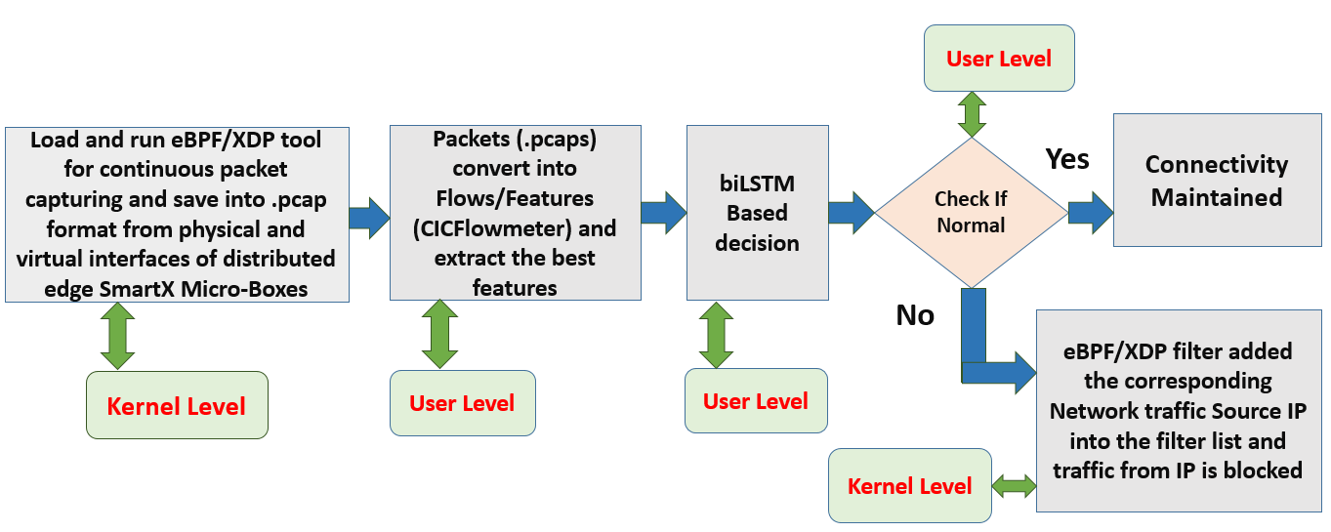}
	\caption{SmartX Intelligent Sec Real Time Implementation Working}
\end{figure}

\begin{figure*}[tp]
	\centering
	\includegraphics[width=120mm, height=6.6cm]{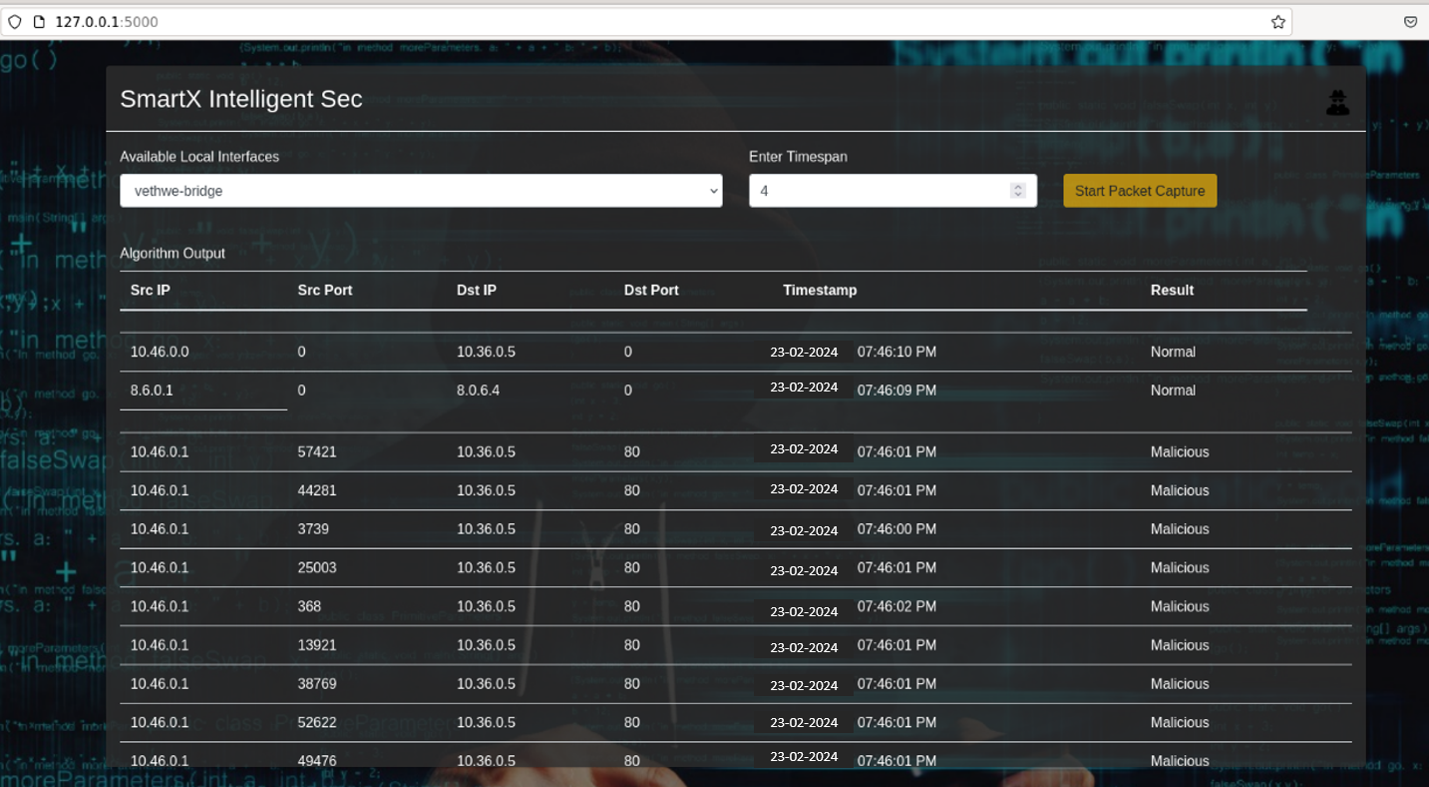}
	\caption{SmartX Intelligent Sec Real Time Implementation}
\end{figure*}

\begin{figure*}[tp]
	\centering
	\includegraphics[width=120mm, height=5.2cm]{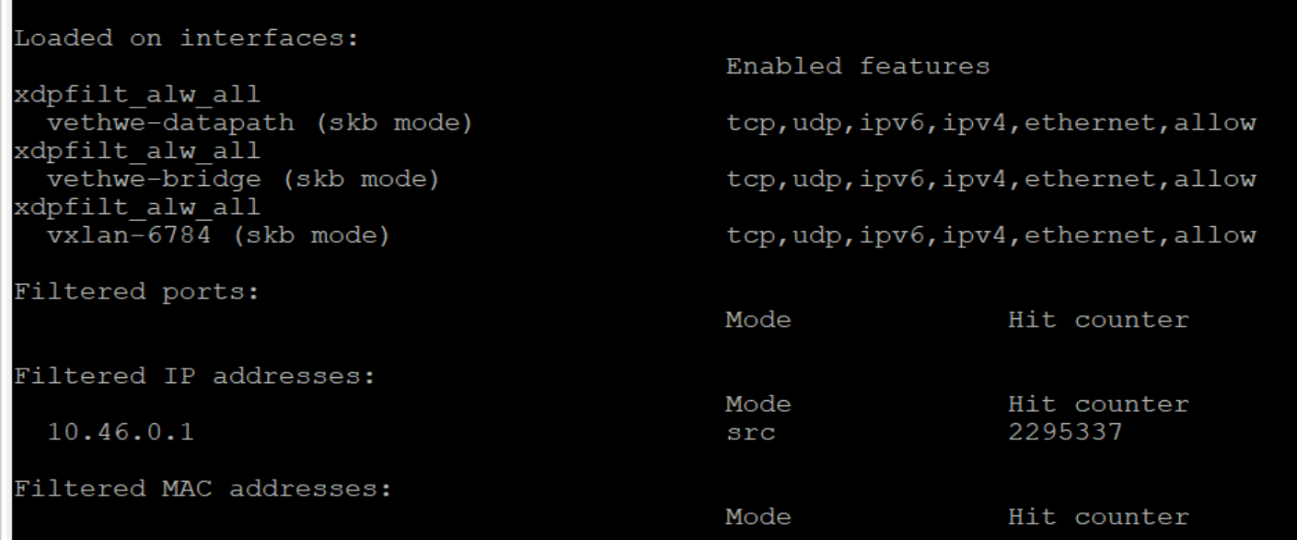}
	\caption{Status of XDP-filter (Attacking Container Pod-IP (10.46.0.1) has been added to filtered IP addresses list)}
\end{figure*}
\subsection{Network Packet Filtering}We utilize the configured XDP tool \cite{bpf}, specifically xdp-filter, for network packet filtering. The eBPF/XDP tool, known for its efficiency, achieves high filter/drop rates, handling tens of millions of packets per second on a single CPU core \cite{bpf}. When the BiLSTM-based network threat detection module identifies incoming traffic as malicious, xdp-filter immediately captures the source IP of the malicious traffic. Consequently, all traffic from that malicious source IP is dropped or filtered, and connectivity with that source IP is terminated.
\subsection{Benefits of using eBPF/XDP}
eBPF/XDP filters out malicious traffic, retrieving only normal packets and thereby significantly mitigating network threats. Modern infrastructures often include numerous virtualized and containerized cloud nodes. Under heavy networking traffic, capturing and filtering packets from these networking ports can consume substantial compute and storage resources, potentially disrupting legitimate services. Therefore, lightweight software functions are essential for managing various networking ports flexibly while minimizing resource consumption. In such scenarios, eBPF/XDP-based packet capturing and filtering of malicious traffic is an optimal solution due to its efficiency and low resource usage.

\section{SmartX Intelligent Sec Prototype Implementation}
Aligned with future Internet testbeds like the Global Environment for Networking Innovation (GENI), the Gwangju Institute of Science and Technology (GIST) has launched the OF@TEIN testbed/playground \cite{oftein, aris,aris2, usman, talia, talia3}. The OF@TEIN playground provides a dynamic array of distributed physical and virtual resources for users and developers to learn operational and development issues and conduct various experiments. It is an overlay, multi-site playground spanning heterogeneous underlay networks, connecting around 14 sites across 10 countries. The playground features "SmartX Micro-Boxes," commodity server-based hyper-converged resources (compute/storage/networking) distributed at multi-site edge locations. These Micro-Boxes support experiments around Cloud computing, Software-defined Networking (SDN), and Network Function Virtualization (NFV). Equipped with cloud-native (containerized) functionalities, the Micro-Boxes function as Kubernetes-orchestrated workers with SDN-coordinated connectivity to other Micro-Boxes. The infrastructure is managed by the "SmartX Playground Tower," which automates the building, operation, and utilization of the playground. The Playground Tower oversees the multi-site playground operation, leveraging several entities, including a Visibility Center, to ensure efficient management.

We implement a real-time prototype of SmartX Intelligent Sec over the OF@TEIN playground, as illustrated in Figure 4 and 5. First, eBPF/XDP captures real-time network traffic from the selected interface (physical or virtual) at the kernel level. Users can specify the packet capturing duration (up to 10 seconds), making the time complexity flexible. For instance, if the user selects 4 seconds, packets are captured for that period, then processed by CICFlowmeter to extract flow features and select the top 20 features. These features are then analyzed by BiLSTM for network threat detection.
If the network traffic is normal, packets continue to move from kernel space to user space. Conversely, if the traffic is identified as malicious, XDP-filter immediately adds the source IP to the filter list, dropping packets from that source at the kernel level with an extremely high drop rate, capable of handling tens of millions of packets per second on a single CPU core.

\textbf{Real Time Testing} We also simulate a DDoS attack, as illustrated in Figure 5. SmartX Intelligent Sec is deployed on the Kubernetes veth interface of the GIST-1 SmartX Edge Micro-Box, part of the OF@TEIN playground. We set the packet capture duration to 4 seconds. During this time, eBPF/XDP captures packets and forwards them to the subsequent modules for further processing, as depicted in Figure 4.

To simulate the DDoS attack, we use a testing machine to launch an attack for 15 seconds on the GIST-1 SmartX Edge Micro-Box using a Docker image of hping3 (TCP SYN flood), with the attacker Pod having the IP 10.46.0.1. When the attacker Pod is activated, eBPF/XDP begins capturing packets for the first 4 seconds. These packets are then displayed as malicious by SmartX Intelligent Sec (detected by BiLSTM).

Subsequently, the eBPF/XDP filter immediately adds the attacker Pod IP 10.46.0.1 to its filter list. As a result, all malicious DDoS traffic from the attacker Pod IP 10.46.0.1 is completely filtered/dropped. The eBPF/XDP filter status in Figure 12 shows that 2,295,337 packets have been filtered/blocked from the attacker Pod IP 10.46.0.1.

\section{Conclusion, Limitations, and Future Work}
We utilize the highly effective BiLSTM deep learning algorithm for network threat detection in conjunction with the fast, efficient, and lightweight eBPF/XDP for network packet capturing and filtering of malicious traffic. This powerful combination addresses contemporary network security challenges robustly. Our real-time prototype demonstrates that SmartX Intelligent Sec provides comprehensive automation features for continuous network packet capturing, network threat detection, and filtering malicious network traffic. However, a limitation of our current work is that the BiLSTM-based network threat detection module operates at the user level. As future work, we plan to shift this module to the kernel space to enhance efficiency further. Additionally, we aim to explore real-time and incremental training methods inside the Linux kernel space.

\section{Acknowledgement}
This work was supported by Institute of Information \& Communications Technology Planning \& Evaluation (IITP) grant funded by the Korea government (MSIT) (No.2019-0-01842, Artificial Intelligence Graduate School Program (GIST)).

\bibliographystyle{ACM-Reference-Format}
\bibliography{samplebase}



\end{document}